\documentclass{aa}  

\usepackage{hyperref}
\usepackage{natbib}
\usepackage{graphicx}
\usepackage{subcaption} 
\usepackage{txfonts}
\begin{document} 
	
	\titlerunning{BNS mergers in Fermi-GBM catalog}
	\title{Identifying potential binary neutron star merger events from the Fermi GBM Gamma-Ray Burst Catalog}

	\author{I. F. Giudice
		\inst{1,2}
		\and
		L. Izzo\inst{2,3}
		\and
		R. Martone\inst{2,4}
		\and
		M. T. Botticella\inst{2}
		\and
		E. Cappellaro\inst{5}
		\and
		R. De Rosa\inst{1,6}
		\and
		M. Della Valle\inst{2,7}
	}
	
	\institute{Università degli Studi di Napoli Federico II, Dipartimento di Fisica "Ettore Pancini"
		\email{ines.giudice@inaf.it}
		\and INAF, Osservatorio Astronomico di Capodimonte, Salita Moiariello 16, I-80131, Naples, Italy 
		\and DARK, Niels Bohr Institute, University of Copenhagen, Jagtvej 128, 2200 Copenhagen, Denmark
		\and Mashfrog Group, Via Giacomo Peroni 400, 00131, Rome, Italy
		\and INAF, Osservatorio Astronomico di Padova, vicolo dell'Osservatorio 5, 35122 Padova, Italy
		\and INFN, Sezione di Napoli, I-80126 Napoli, Italy
		\and ICRANet, Piazza della Repubblica 10, I-65122 Pescara, Italy
	}
	
	
	
	\abstract
	{Short gamma-ray bursts  are expected to be associated with compact object mergers, such as binary neutron star or neutron star-black hole systems, and are key high-energy  multimessenger events. The detection of GRB 170817A, coinciding with the gravitational wave signal GW170817 from a BNS merger, confirmed the link between sGRBs and compact object mergers. Similarly, GRB 150101B displayed remarkable similarities to GRB 170817A, further supporting its association with compact binary mergers. }
	{The objective of this study is to uncover the intrinsic properties that differentiate merger-associated sGRBs from other GRBs by analyzing the  Fermi GBM Burst Catalog and using GRB 170817A and GRB 150101B as reference events, enhancing our ability to select events from this class  and promptly to search for their electromagnetic counterpart. }
	{We employed a clustering technique to classify GRBs based on their observed properties in gamma-rays ($T_{90}$, E$_{peak}$ and fluence). Prior to clustering, we tested three dimensionality reduction techniques, among which Uniform Manifold Approximation and Projection demonstrated the best performance making it the preferred technique for our analysis. This combination of dimensionality reduction and clustering analysis allowed us to group GRBs with similar characteristics, with a focus on identifying those most likely associated with BNS mergers. }
	{Our analysis successfully identified a cluster of sGRBs events with characteristics consistent with sGRB merger-associated. A comparison between our sample of candidates and known kilonova candidates associated with sGRBs, identified through other methodologies, further validated our approach.
	}
	{}
	
	
	\maketitle
	%
	
	\section{Introduction}   
	The advent of multimessenger astronomy is transforming our understanding of gamma-ray bursts (GRBs), particularly those linked to compact object mergers such as binary neutron stars (BNSs) or neutron star-black hole (NS-BH) systems. These catastrophic events produce high-energy gamma-ray emissions and gravitational waves (GWs), enabling detections across multiple observational channels. The detection of GRB 170817A \citep{goldstein2017ordinary}, coincident with the GW signal GW170817 \citep{abbott2017gw170817}, marked a milestone in multimessenger astronomy. This event not only confirmed the association between short gamma-ray bursts (sGRBs) and compact object mergers but also revealed an optical kilonova (KN) counterpart, AT2017gfo, powered by the radioactive decay of r-process elements synthesized in the merger ejecta \citep{smartt2017kilonova,kasen2017origin,pian2017spectroscopic}.
	
	The sGRBs are characterized by ultrarelativistic, highly collimated jets with Lorentz factors of 100–1000 and prompt gamma-ray emissions lasting under two seconds \citep{kou1993}. These jets are typically observed along the axis of the burst, known as the on-axis view, which results in a highly focused and intense emission. Off-axis GRBs, on the other hand, occur when the relativistic jet is observed from an angle outside its core, leading to reduced beaming effects and a diminished observed flux. The Doppler boosting, which typically amplifies emission for on-axis observations, becomes less effective as the viewing angle increases \citep{salafia2016light}. This reduction in the effective Lorentz factor shifts the observed peak energy to lower frequencies and softens the spectrum, making off-axis GRBs appear fainter and more challenging to detect \citep{gill2020linear,lamb2017electromagnetic,lazzati2018late}. Additionally, slightly off-axis GRBs typically exhibit longer apparent durations compared to on-axis events, as the emission from the outer regions of the jet, including cocoon emission, reaches the observer later in time. This broader angular distribution, along with variations in photon arrival times, leads to extended signals \citep{chakyar2025effect,salafia2016light}, providing invaluable insights into jet structures and energy distributions \citep{lazzati2017off}.
	In the context of compact binary mergers, off-axis GRBs may play a crucial role, as they are expected to outnumber on-axis events due to their wider angular distribution \citep{fong2015decade}. However, they remain often undetected due to large error regions in their localization, complicating follow-up searches for optical counterparts. GRB 170817A exemplified an off-axis GRB, where the relativistic jet was observed at an angle away from its core, leading to fainter prompt gamma-ray emission. Despite this, the event highlighted the potential for identifying similar cases that might otherwise be missed due to jet misalignment \citep{goldstein2017ordinary,lazzati2018late}. Its simultaneous detection of gamma rays, electromagnetic counterpart and GWs reinforced the value of multimessenger observations. Another notable case, GRB 150101B, exhibited features akin to GRB 170817A, reinforcing the link between sGRBs and compact mergers even without a GW counterpart. Observed in a luminous elliptical galaxy with no star formation, it displayed a bright optical counterpart consistent with a luminous KN and an extended X-ray afterglow \citep{troja2018luminous}.The development of methodologies to identify these elusive off-axis events is essential for the full characterization of the GRB population resulting from compact binary mergers and for facilitating the detection of their KN counterparts, a critical component of multimessenger follow-up.
	
	This study builds on the multimessenger framework with the aim of improving the identification of potential electromagnetic counterparts to compact object mergers through the application of machine learning techniques. The objective of this approach is to facilitate the identification of GRB candidates, particularly those detected by the same detector or instrument, that are likely to result from compact binary mergers. We employed a clustering algorithm to isolate GRBs with characteristics similar to known KN-associated events, such as GRB 150101B \citep{troja2018luminous} and GRB 170817A \citep{goldstein2017ordinary}, by analyzing the Fermi Gamma-Ray Burst Monitor (GBM) Catalog. While purely statistical, this approach can help identify GRBs with similar intrinsic properties, potentially highlighting events that share characteristics with the known off-axis sGRBs from binary neutron star mergers, which features are reported in Table \ref{refgrbs}. 
	
	\begin{table}[h!]
		\centering
		\caption{Properties of GRBs with confirmed association to a KN.}
		\label{refgrbs}
		\begin{tabular}{cccc} 
			\hline
			\hline
			& fluence  & E$_{peak}$ &  T$_{90}$ \\
			&\textnormal{($10^{-7}$ erg/cm$^{2}$)} &(\textnormal{keV}) &(\textnormal{s})\\
			\hline
			170817A &2.79$\pm{0.17}$ &  215$\pm{56}$ &  2.48$\pm{0.47}$ \\
			150101B & 0.76 $\pm{0.11}$ &208$\pm{109}$  &0.48$\pm{0.10}$ \\
			\hline
		\end{tabular}
	\end{table}
	
Since identifying off-axis GRBs based only on prompt emission is challenging, given that the off-axis nature is typically inferred from afterglow behavior, the clustering algorithm does not explicitly distinguish between on- and off-axis events. Our approach focuses on identifying events in the Fermi dataset that exhibit prompt emission properties similar to the two known GRBs associated with KNe. The goal is to develop an algorithm capable of flagging or sending alerts for events detected by Fermi that may be indicative of short GRBs viewed off-axis, thus facilitating follow-up observations and improving the likelihood of detecting electromagnetic counterparts to compact binary mergers. This prompt-based methodology provides a rapid tool for early classification of the most promising GRB candidates for follow-up that might otherwise be overlooked in gamma-ray observations. Specifically, it can determine whether a newly detected GRB falls within the cluster of interest in under two minutes, making it well suited for real-time follow-up prioritization. In addition, a pipeline has been implemented for associating host galaxies for potential nearby GRB events. By refining our ability to pinpoint host galaxies of nearby sGRBs, we can improve the precision of localization and the prompt detection of KNe. 
	Both the clustering methodology and the host galaxy association pipeline have been tailored for events detected by the Fermi GBM, ensuring methodological consistency and robustness. By focusing on a single instrument’s observations, our approach mitigates potential biases introduced by data from multiple detectors, ultimately increasing the reliability and precision of our results in identifying sGRBs with characteristics indicative of compact binary mergers.
	
	The paper is organized as follows: Section \ref{sec:data} describes the dataset utilized in this study, while Section \ref{sec:method} details the applied methodology. The results of the clustering analysis are presented in Section \ref{sec:results}, and the properties of the merger-associated sub-sample are discussed in Section \ref{sec:merger-GRB}. Section \ref{sec:hostpipeline} provides a broader discussion on the host galaxy association pipeline.  A general discussion of the results, including a comparison with previous studies, is presented in Section \ref{discussion}. Finally, Section \ref{sec:conclusion} summarizes the key findings and conclusions.

	\section{Data sample}
	\label{sec:data}
	Our methodology was applied to the Fermi GBM Burst Catalog \citep{von2020fourth,gruber2014fermi,von2014second,bhat2016third}, a comprehensive dataset that collects GRBs detected by the GBM on board the Fermi spacecraft. Fermi-GBM consists of two main detectors: two BGOs, covering the high energy range from 200 keV up to 40 MeV, and twelve NaI scintillation detectors, sensitive to low-energy GRB and covering the range from 8 keV to $\sim$ 1 MeV\citep{meegan2009fermi,connaughton2015localization}. The nearly continuous monitoring of the sky and the sensitivity to a broad range of GRBs make the Fermi GBM a suitable choice for our goal of identifying KN-associated events.  Since the beginning of Fermi operations in 2008, over 3000 GRB bursts have been recorded, each systematically classified and cataloged for further study\footnote{The Fermi GBM burst catalog can be queried at the following link \url{https://heasarc.gsfc.nasa.gov/w3browse/fermi/fermigbrst.html}}. Among the parameters included, we specifically selected  fluence, E$_{peak}$, and T$_{90}$, as these are the only parameters in the Fermi Burst Catalog that are directly related to the intrinsic properties of an event, encapsulating information about the energy release and temporal profile of GRBs. The  fluence represents the total energy emitted by the burst per unit area, integrated over the entire duration of the event and across the detector’s energy band. The $T_{90}$ parameter is defined as the time interval during which 90\% of the burst's total fluence is accumulated. Finally, we utilized the peak energy $E_{peak}$ derived from the Band model. While these parameters are not sufficient to distinguish between different progenitor types (e.g., BNS, NSBH, or NSWD mergers), they provide a robust framework for characterizing each burst, thereby facilitating  the identification of subgroups with similar prompt emission properties. 
	
	Since our goal is to identify clusters within the GRB population, we included all events, regardless of duration, in our clustering analysis. This approach captures underlying patterns in the data, thereby facilitating a robust identification of relevant subpopulations. Analyzing both long and short GRBs ensures comprehensive coverage, offering insights into global GRB properties while refining the search for sGRB candidates as potential optical counterparts to GW events.
	
	The data was retrieved from the catalog in January 2025. While the catalog contains over 3917 GRB events, our analysis was limited to bursts with available measurements for fluence, E$_{peak}$, and T$_{90}$. Applying this restriction reduced the dataset to a subset of 3585 GRBs. 
	
	\section{Methodology}
	\label{sec:method}
	In this section, we outline the methodology designed to identify subpopulations within the Fermi GBM catalog. Data pre-processing (Section \ref{sec:pre}) is an essential first step in this process, as it mitigates the impact of extreme values. We then applied dimensionality reduction techniques, which serve to highlight key features in the dataset. The three different techniques tested are discussed in this section \ref{sec:dimensionalityred}. Following this, clustering was performed, applying the algorithm introduced in Section \ref{sec:kmeans} on all the dimensionality reduction techniques tested.  Finally, Section \ref{sec:ncluster} details the strategy for selecting the optimal number of clusters $n_{cluster}$ and dimensionality reduction technique. This process includes metric-based selections followed by additional computations to evaluate the quality of the chosen $n_{cluster}$. 
	
	\subsection{Pre-processing phase}
	\label{sec:pre}
	
	Given that our selected parameters span a broad range of values, initial pre-processing is crucial for effective clustering. To reduce the impact of extreme values and outliers, we applied a logarithmic transformation to the raw data using the {\tt numpy} Python module. Following this, we standardized the data to ensure that each feature contributed equally to the clustering process.  Standardization is a critical pre-processing step, especially for distance-based clustering algorithms like those relying on Euclidean distance, which are sensitive to the scale of input features. Without normalization, features with inherently larger scales could dominate those with smaller ones, distorting the clustering results. To achieve this, we used the {\tt StandardScaler} class implemented in the {\tt sklearn.preprocessing} module in Python. The StandardScaler employs the Z-score formula \citep{mohamad2013research}, in which, given a raw set of data for each feature, the Z-score for each data point  $x_{ij}$ is calculated as:
	
	\begin{equation}
		x_{ij}=Z(z_{ij})=\frac{x_{ij}-\bar{x}_{j}}{\sigma_{j}},
	\end{equation}
	where $\bar{x_{j}}$ and $\sigma_{j}$ are the sample mean and standard deviation of the $j$ feature, respectively, and $i$ indexes the individual data points within that feature.  This transformation centers each feature around a mean of zero with unit variance, ensuring both the magnitude and variability of features are equalized. The {\tt fit\_transform} method efficiently computes these statistics and applies the standardization, thereby enhancing feature comparability and improving the reliability of the clustering process.

	\subsection{Dimensionality reduction techniques}
	\label{sec:dimensionalityred}
	Dimensionality reduction techniques are essential for simplifying complex, high-dimensional datasets, particularly in the context of clustering and visualization. In this study, we applied three complementary methods: principal component analysis (PCA), t-distributed stochastic neighbor embedding (t-SNE), and uniform manifold approximation and projection (UMAP). While all three techniques aim to reduce dimensionality, they differ fundamentally in their approach. The PCA is a linear method that focuses on preserving global variance, making it well suited for datasets with predominantly linear structures. In contrast, t-SNE and UMAP are nonlinear methods, excelling at capturing local relationships and uncovering latent patterns in data with complex, nonlinear structures. By leveraging the strengths of these diverse techniques, we gain complementary insights into the data, enabling both robust clustering and detailed visualization. 
	In all the approaches the parameter {\tt n\_components} was set to 2, thereby reducing the data to two dimensions and facilitating both the clear visualization of the dataset's structure and  the highlighting of its most relevant features.  Additionally, the methodology presented in this section was performed without applying dimensionality reduction to assess its impact on clustering performance. The results are discussed in Appendix \ref{sec:nored}.
	
	\subsubsection{PCA} 
	
	The PCA is a linear dimensionality reduction technique that seeks to identify the directions (principal components) along which the variance in a dataset is maximized \citep{mackiewicz1993principal}. The technique transforms high-dimensional data into a set of orthogonal components, ranked by their variance, where the first component captures the most variance, the second the next highest, and so on. The method first computes the covariance matrix on standardized data, and performs eigenvalue decomposition to extract principal components. The data is then projected onto a subset of these components to reduce dimensionality while preserving most of the variance.  For the purposes of this study, the PCA class implemented in the {\tt sklearn.decomposition} module in Python was used. 
	
	\subsubsection{t-SNE} 
	
	The t-SNE is a nonlinear dimensionality reduction technique \citep{van2008visualizing} that maps high-dimensional data into a lower-dimensional space—typically two or three dimensions—where clusters and patterns become visually interpretable. The method first computes pairwise similarities between data points in the high-dimensional space, with perplexity serving as a critical parameter that balances the representation of local and global structures, thereby influencing the scale of patterns captured. These similarities are subsequently projected into a lower-dimensional space using a t-distribution, which preserves local relationships while maintaining separation between dissimilar points. The projection is refined iteratively through gradient descent, minimizing the Kullback-Leibler divergence between the high- and low-dimensional representations, ultimately revealing latent structures within the data.
	
	For our implementation, we used the {\tt TSNE} class implemented in the {\tt sklearn.manifold} module in Python. The perplexity parameter was configured to a value of 30, which balances local and global structure representation, though values generally range from 5 to 50 depending on dataset size and density. Finally the {\tt fit\_transform} was applied to simultaneously compute and perform the dimensionality reduction.

	\subsubsection{UMAP} 
	
	The UMAP is a nonlinear dimensionality reduction technique designed to map high-dimensional data to a lower-dimensional space while preserving both local and global data structures \citep{mcinnes2018umap}.
	The algorithm constructs a weighted graph of nearest neighbors in the high-dimensional space, with the {\tt n\_neighbors} parameter controlling the balance between local and global data preservation. Optimization is then performed using stochastic gradient descent to minimize a cross-entropy loss function, aligning the high-dimensional relationships with their low-dimensional representations.
	In our analysis, we employed the {\tt UMAP} class implemented in the {\tt umap}- {\tt learn} library in Python with {\tt n\_neighbors=15}, {\tt min\_dist=0.1} and {\tt  metric='euclidean'}. 
	
	\subsection{K-Means clustering algorithm}
	\label{sec:kmeans}
	
	We employed K-Means, an unsupervised machine learning clustering algorithm, to uncover patters within our pre-processed Fermi GBM catalog. The K-Means algorithm operates by assigning each data point to the nearest cluster centroid, thereby minimizing the within-cluster sum of squares (WCSS). In this context, machine learning serves as a powerful tool for pattern recognition, allowing us to classify GRB candidates into distinct groups based on their similarities. Similar methodologies have been successfully applied in astrophysical data analysis, such as the classification of GRBs based on their physical properties \citep{chen2025classification}. The algorithm is particularly effective for identifying underlying structures in complex datasets, as it enables the segmentation of the data into $k$ clusters based on shared observational characteristics  fluence,  T$_{90}$, and E$_{peak}$ in this analysis. 
	
	We implemented K-Means using the {\tt K-Means} module from the {\tt sklearn.cluster} package.  This method selects initial centroids randomly from the dataset, which, while straightforward, may introduce variability in the clustering results. However, to counteract this, we conducted multiple iterations  {\tt max\_iter}=100000 with different random initializations to enhance the robustness and stability of the clustering outcomes. After the initial selection of centroids, K-Means iteratively minimizes WCSS through a process of assigning data points to the nearest centroid and recalculating centroid positions until the cluster assignments stabilize \citep{lloyd1982least}. This iterative refinement allows for effective clustering, even when starting from random points.

	\subsection{Cluster number selection}
	\label{sec:ncluster}
	\begin{figure*}[h]
		\centering
		\begin{subfigure}[t]{0.32\textwidth}
			\centering
			\includegraphics[width=\textwidth]{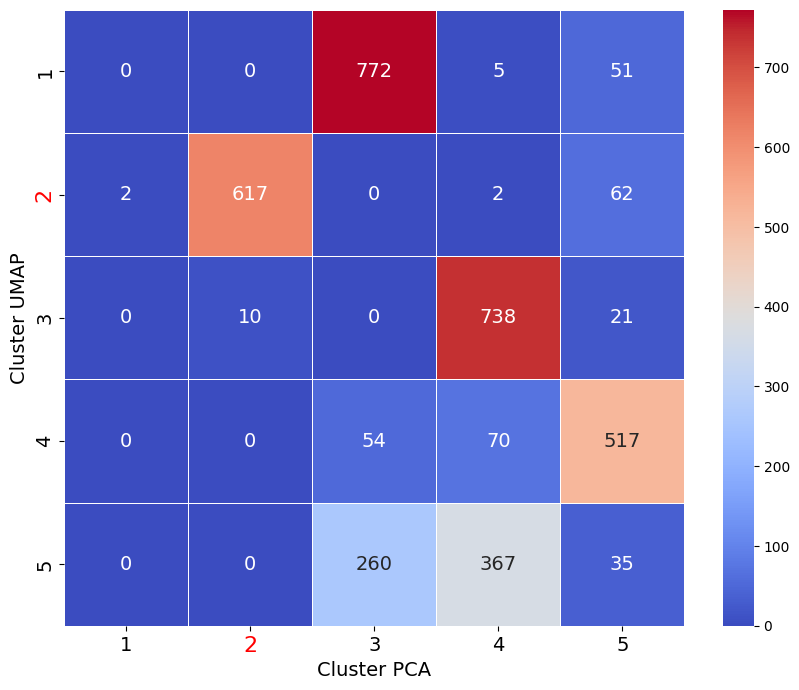}
			\caption{Cluster overlap between PCA and UMAP.}
			\label{fig:heatmap1}
		\end{subfigure}
		\hfill
		\begin{subfigure}[t]{0.32\textwidth}
			\centering
			\includegraphics[width=\textwidth]{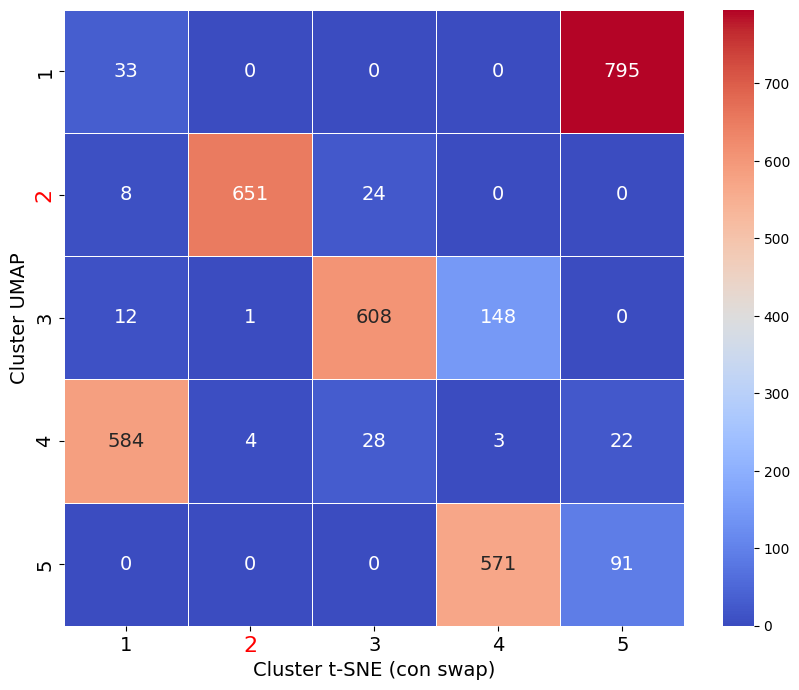}
			\caption{Cluster overlap between UMAP and t-SNE.}
			\label{fig:heatmap2}
		\end{subfigure}
		\hfill
		\begin{subfigure}[t]{0.32\textwidth}
			\centering
			\includegraphics[width=\textwidth]{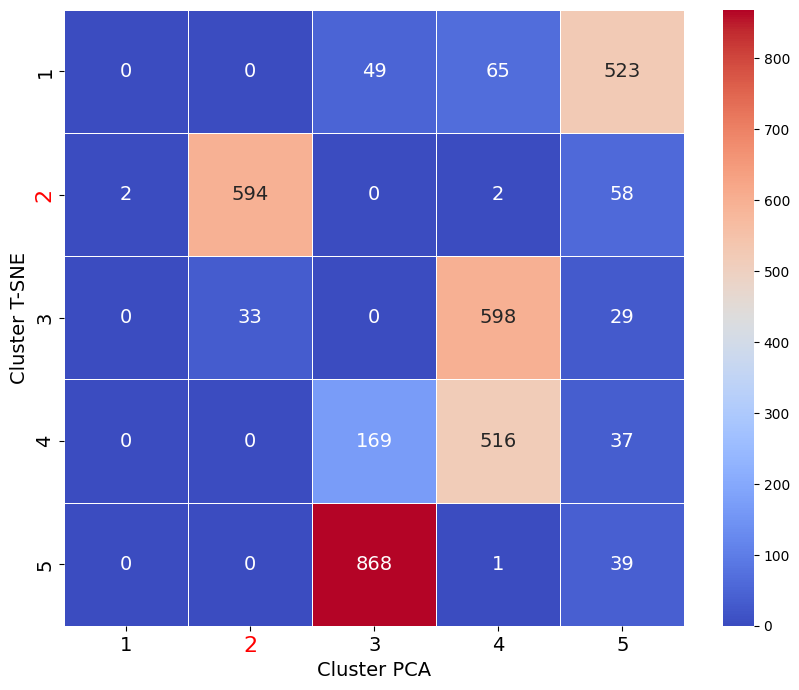}
			\caption{Cluster overlap between PCA and t-SNE.}
			\label{fig:heatmap3}
		\end{subfigure}
		\caption{Cluster overlap across the three dimensionality reduction techniques. The label of the clusters of interest are highlighted in red. }
		\label{fig:combinedheatmaps}
	\end{figure*}
	
	The selection of the optimal number of clusters is a critical step in any clustering analysis, as it directly influences the interpretability and reliability of the results.  K-Means does not automatically determine the optimal number of clusters $n_{cluster}$ and to address this issue we compared the clustering quality across the three different dimensionality reduction techniques (PCA, t-SNE, and UMAP) using the Silhouette score and the Davies–Bouldin index (DBI) for values of $n_{cluster}$ ranging from 2 to 8. Both metrics provide complementary insights into the clustering performance. The Silhouette method is a widely used technique for evaluating the consistency of clustering by assessing the quality of different $n_{cluster}$ configurations \citep{rousseeuw1987silhouettes}. This metric quantifies the similarity of each data point to its designated cluster relative to neighboring clusters, thereby evaluating the distinctiveness and separation between the various clusters in the dataset. The resulting Silhouette score provides a measure of cluster quality, ranging from -1, indicating poor separation, to 1, which signifies well defined clusters. This method is also particularly useful for determining the optimal number of clusters, as it reveals how compact and distinct each grouping is. Then another parameter that evaluates the goodness of the clustering is the already mentioned DBI \citep{davies1979cluster}. The DBI quantifies the average similarity ratio of each cluster with its most similar cluster, effectively measuring how well separated and compact the clusters are. Specifically, for each cluster, the DBI is calculated as the ratio of the sum of intra-cluster distances (how dispersed the points are within the same cluster) to the inter-cluster distances (the distance between cluster centroids). A lower DBI, $\leq 1$, indicates better clustering performance, suggesting that clusters are not only well separated but also cohesive within themselves, while higher scores suggest overlapping clusters.
	The clustering computation through Silhouette and DBI have been performed with the {\tt silhouette\_score} and the {\tt davies\_bouldin\_score} modules from {\tt sklearn.metrics}.
	
	\begin{table}[h!]
		\centering
		\caption{Comparison of clustering scores across PCA, t-SNE, and UMAP.}
		\label{tab:comparison_tot}
		\begin{tabular}{lccc} 
			\hline
			\hline
			Method & DBI & Silhouette & $n\_cluster$ \\
			\hline
			t-SNE & 1.00 & 0.39 & 2 \\
			t-SNE & 0.84 & 0.42 & 3 \\
			t-SNE & 0.79 & 0.42 & 4 \\
			t-SNE & 0.79 & 0.42 & 5 \\
			t-SNE & 0.83 & 0.40 & 6 \\
			t-SNE & 0.83 & 0.41 & 7 \\
			t-SNE & 0.83 & 0.40 & 8 \\
			\hline
			PCA & 0.83 & 0.39 & 2 \\
			PCA & 0.92 & 0.37 & 3 \\
			PCA & 0.72 & 0.37 & 4 \\
			PCA & 0.74 & 0.36 & 5 \\
			PCA & 0.75 & 0.36 & 6 \\
			PCA & 0.82 & 0.33 & 7 \\
			PCA & 0.83 & 0.33 & 8 \\
			\hline
			UMAP & 0.88 & 0.44 & 2 \\
			UMAP & 0.82 & 0.44& 3 \\
			UMAP & 0.83 & 0.42 & 4 \\
			UMAP & 0.72 & 0.45 & 5 \\
			UMAP & 0.76 & 0.43 & 6 \\
			UMAP & 0.78 & 0.42 & 7 \\
			UMAP & 0.80 & 0.42 & 8 \\
			\hline
		\end{tabular}
	\end{table}

	The lower end of the tested range, specifically $n_{cluster}$ = 2 and 3, given the specific interest in GRBs with merger-like properties, might be too simplistic to capture the nuanced differences among merger-related events. Higher values, such as $n_{cluster}$ = 8, were considered to account for potential substructures within the data, which might reveal smaller, distinct subgroups that could reflect subtle variations in properties such as host galaxy characteristics or physical offsets. The Silhouette and DBI metrics, when accounting for these previous considerations, indicated that configurations with $n_{cluster}$=4 or $n_{cluster}$=5 (see Table \ref{tab:comparison_tot}) achieved relatively high-quality clustering. However, the final choice of $n_{cluster}$=5 was motivated by further analysis. With $n_{cluster}$=5, the clusters became more distinct and cohesive, making it easier to interpret and assign data points to specific groups. Importantly, choosing $n_{cluster}$=5 does not result in the loss of information. In fact, the cluster containing the two reference GRBs (GRB 170817A and GRB 150101B) remained consistently grouped together in the most effective combination (UMAP associated with K-Means), with both $n_{cluster}$=4 and $n_{cluster}$=5, ensuring the preservation of key scientific insights.
	
	\subsection{Comparison of results from different methods}
	
	Following the selection of $n_{cluster}$, the consistency of the outcomes from the three dimensionality reduction techniques applied to K-Means was investigated. For this purpose,  adjusted rand index (ARI) and normalized mutual information (NMI) were employed, as these metrics evaluate the alignment and consistency between clustering results derived from different algorithms. Specifically, ARI quantifies the similarity between clusterings by evaluating the number of data points that are correctly grouped. This metric ranges from -1 to 1, with 1 indicating a higher degree of alignment. The NMI measures the shared information between two clustering and ranges from 0 to 1. A score near 1 defines a high degree of similarity between the clustering, reflecting well defined clusters, whereas a value close to 0 suggests minimal agreement. 
	The results (see Table \ref{tab:comparisonarinmi5}) revealed moderate-to-high agreement among the methods, reflecting the varying nature of each approach. Interestingly, the highest agreement was observed between UMAP and t-SNE, as both techniques aim to preserve the local structure of the data while reducing its dimensionality. In this analysis the {\tt adjusted\_rand\_score} and {\tt normalized\_mutual\_info\_score} modules from the {\tt sklearn\_metrics} Python library were employed.

	\begin{table}[h!]
		\centering
		\caption{Comparison of clustering scores for $n_{cluster}$ =  5 across PCA, t-SNE, and UMAP combinations.}
		\label{tab:comparisonarinmi5}
		\begin{tabular}{lcc} 
			\hline
			\hline
			Method & ARI & NMI  \\
			\hline
			
			t-SNE - UMAP  & 0.86 & 0.84  \\
			t-SNE - PCA & 0.55 & 0.61  \\
			PCA - UMAP & 0.55& 0.61 \\
			
			\hline
		\end{tabular}
	\end{table}
	
	Additionally, ARI and NMI were also computed to compare the clustering results within K-Means and DBSCAN. DBSCAN (Density-based spatial clustering of applications with noise) is a density-based clustering algorithm that groups points based on their proximity and density. While this algorithm was not part of our GRB clustering analysis, the score computation was conducted to further reinforce our selection of $n_{cluster}$=5 by demonstrating agreement within different methodologies. The results are discussed in Appendix \ref{sec:dbscan}.

	In order to further analyze the consistency and overlap between clusters generated by the three different dimensionality reduction approaches, heatmaps were computed for $n_{clusters}$=5. These heatmaps (see Figure \ref{fig:combinedheatmaps}) illustrate the degree of overlap between clusters obtained through PCA, t-SNE, and UMAP. The results indicated a high degree of agreement in the clustering assignments, particularly for the cluster of primary interest. This cluster consistently exhibits strong overlap across the three methods, thereby reinforcing the reliability of the clustering results and providing confidence in the interpretation of this subgroup as a distinct population.
	
	\section{Clustering results}
	\label{sec:results}

\begin{figure}[!ht]
		\centering
		\begin{minipage}[t]{0.5\textwidth}
			\centering\
			\includegraphics[width=\textwidth]{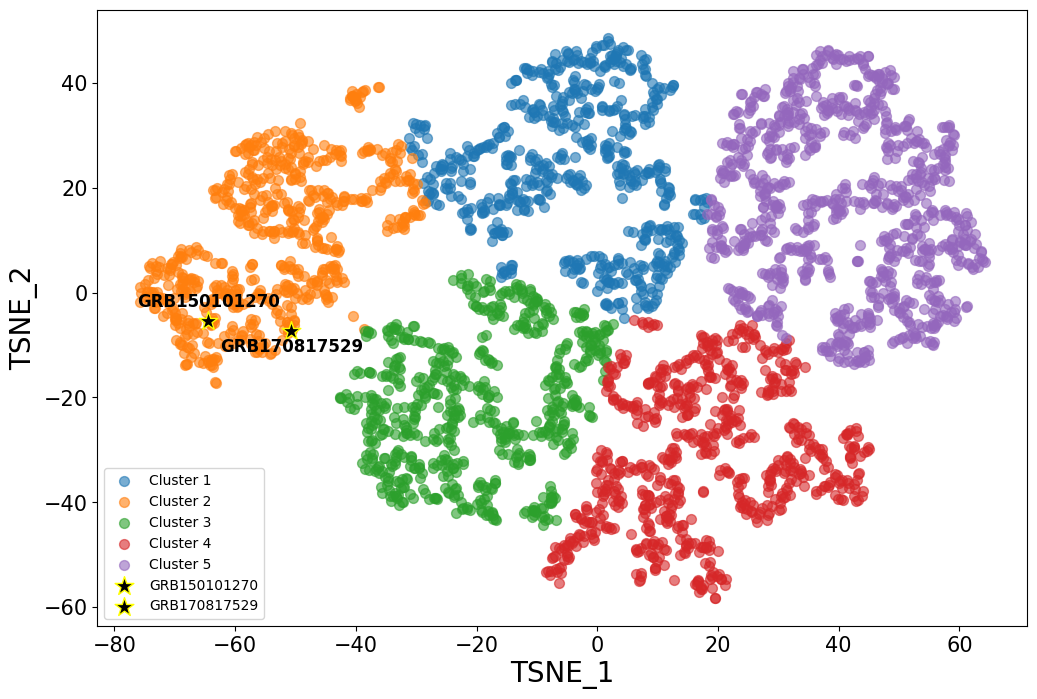}
			\includegraphics[width=\textwidth]{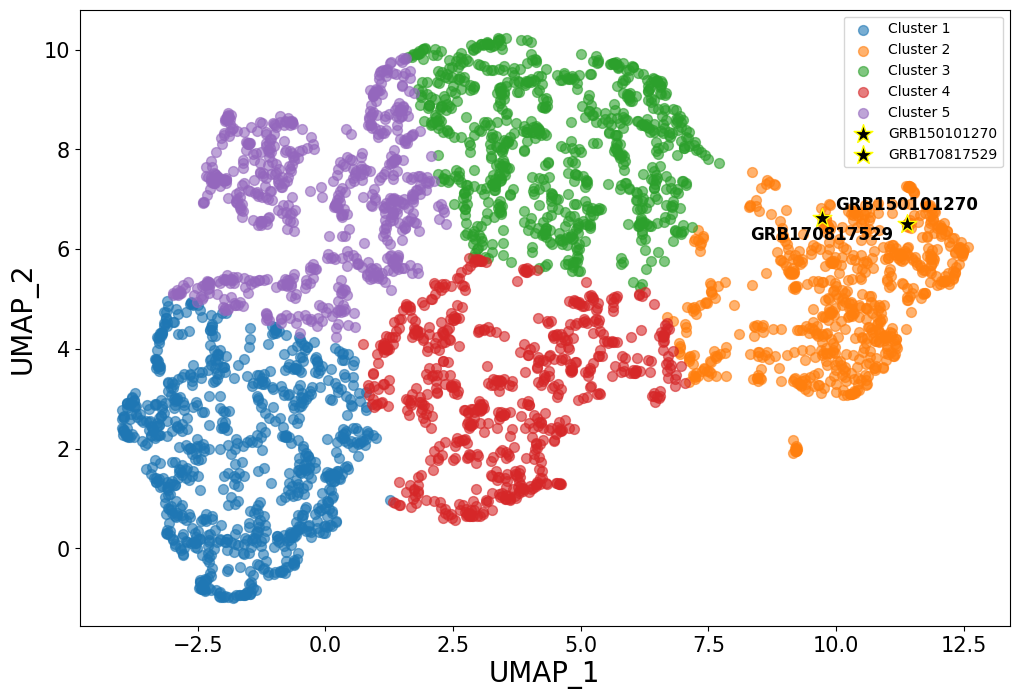}
			\includegraphics[width=\textwidth]{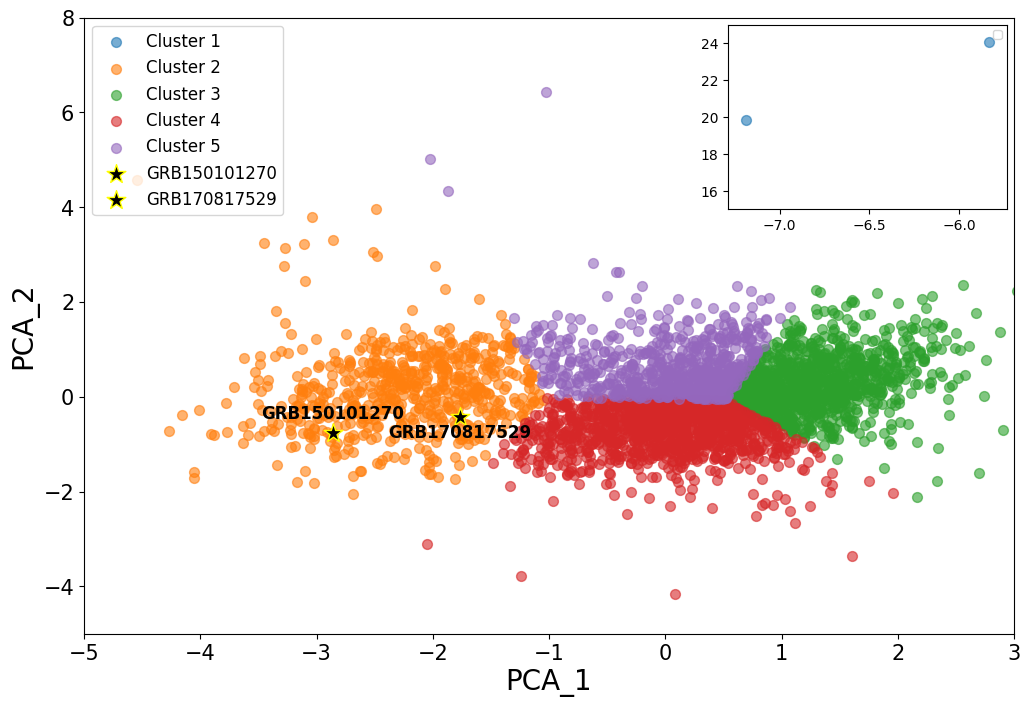}
			\caption{From top to bottom, t-SNE-based, UMAP-based and PCA-based  results for the n=5 selection clustering. . }
			\label{fig:clustering}
		\end{minipage}
		\hfill
	\end{figure}

	Based on the evaluation of clustering metrics, we applied our clustering algorithm to the Fermi GBM dataset with $n_{cluster}$=5 for all three dimensionality reduction methods, revealing distinct subgroups within the GRB dataset. Each cluster, illustrated in Figure \ref{fig:clustering}, represents a statistical division of the GRB sample based on machine learning techniques and is characterized by similar values of T${90}$, fluence, and E $_{peak}$. The full parameter distributions for each cluster are shown in Appendix \ref{distr}.
	
	A notable outcome of the clustering we obtained is the consistent grouping of GRB 150101B and GRB 170817A across all three dimensionality reduction methods, underscoring their shared physical characteristics. In Figure \ref{fig:clustering} these two GRBs are prominently marked as black stars to highlight their proximity within the identified clusters. As indicated by the superior Silhouette and DBI scores (see Table \ref{tab:comparison_tot}), UMAP emerged as the most effective dimensionality reduction technique for this analysis. Therefore, from this point forward, we primarily reference results derived from this method. 
	
\begin{figure}[h]
		\centering
		\includegraphics[width=\hsize]{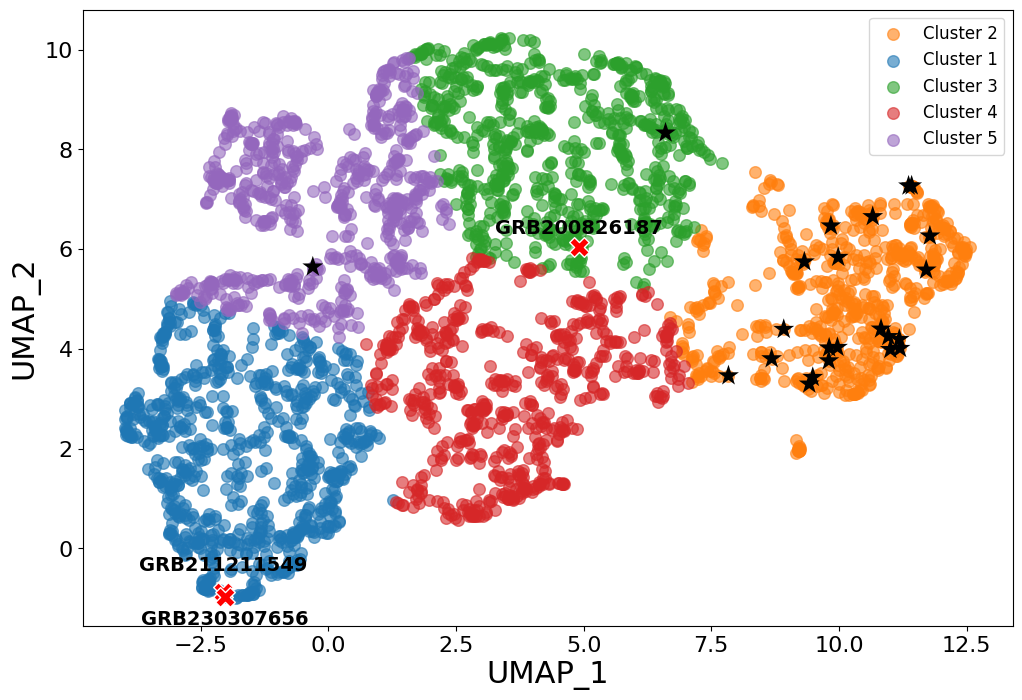}
		\caption{Comparison between our clustering results and the sGRBs from \citet{fong2022short} catalog, with sGRBs marked as black stars.
			Additionally, the plot highlights in red the GRB200826A associated with a collapsar and the two long GRBs associated with KNe (GRB211211A and GRB230307A). The discussion on the long kN-associated GRBs can be found in Section \ref{discussion}.}
		\label{comparisonfong}
	\end{figure}

	Among the five clusters obtained with UMAP dimensionality reduction, Cluster 2 likely identifies a subpopulation of potential KN candidates, as it includes both reference GRBs, suggesting it encompasses GRBs with intrinsic properties similar to those of merger-associated events. This reinforces the robustness of our clustering approach and highlights Cluster 2 as a promising subgroup for further investigation.
	The properties of Cluster 2 are detailed in Table \ref{cluster1} and the centroids of the five clusters are summarized in Table \ref{centroids}. To facilitate interpretation, we applied an inverse transformation to map the centroids back to the original 3D space, allowing us to recover the corresponding values of the main observable features.  Notably, the centroid of Cluster 2 exhibits a clear alignment with the sGRB class  while the other clusters exhibit features close to long GBRs. 
	Moreover, the high  E$_{peak}$ centroid of Cluster 2 may reflect the presence of both on-axis and off-axis events. Several GRBs within the cluster exhibit E$_{peak}$ values consistent with off-axis sGRBs, such as GRB 170817A, 150101B, and 160821B, which show comparatively lower  E$_{peak}$ values. At the same time, the cluster includes very energetic events like GRB 090510. This suggests that the centroid value represents an average across a broad distribution and does not exclude the presence of off-axis sGRBs within the cluster. Therefore, it remains reasonable to assume that such events are part of this population. These findings also suggest that, in our analysis, T$_{90}$ is likely the key parameter driving the classification of GRBs as merger-like events versus collapsars. This interpretation is further supported by the exclusion of GRB 200826A from our merger-like cluster (see Figure \ref{comparisonfong}). Although its rest-frame duration (~0.5 s) places it within the short GRB category, GRB 200826A exhibited several features typical of collapsars \citep{rossi2022peculiar, wang2022grb}: it was energetically soft, followed the Amati relation, and was accompanied by an optical/NIR bump consistent with a supernova. 
	
	To strengthen our association of this cluster with sGRBs, we investigated the sGRB catalog by \citep{fong2022short}. By narrowing our analysis to the events detected by Fermi, we identified the majority of their sGRBs within our cluster. 
	As illustrated in Figure \ref{comparisonfong} a few elements of their sample are distributed in other clusters. In particular, within Cluster 1 we found GRB 211211A which has been linked to a KN (see Section \ref{discussion} for further details).

	The other two GRBs located outside our cluster, GRB 170728B (Cluster 3) and GRB 151229A (Cluster 5), exhibit an effective observed duration in the Fermi GBM catalog that is longer than typical short GRBs. These GRBs have  been considered as sGRBs in the work of \citep{fong2022short}, given that their T90 observed duration is shorter than 2s as reported by the Swift/BAT detector \citep{ukwatta2017grb, lien2015grb}. 
	
\begin{table}[h]
	\centering
	\caption{Centroids properties in the original three-dimensional space. $n$ represents the number of elements in each cluster. }
	\label{centroids}
	\begin{tabular}{ccccc}
		\hline
		\hline
		& Fluence & E$_{peak}$ &  T$_{90}$ & $n$ \\
		& $(10^{-6}$ \textnormal{erg/cm$^{2})$} & \textnormal{(keV)} & \textnormal{(s)} & \\
		\hline
		
		Cluster 1 & 19.75 & 186.60 & 51.72 & 914 \\
		Cluster 2 & 0.40  & 596.53 & 0.79  & 657 \\
		Cluster 3 & 1.19  & 86.78  & 8.34  & 695 \\
		Cluster 4 & 3.87  & 380.57 & 24.83 & 621 \\
		Cluster 5 & 3.92  & 65.29  & 36.00 & 698 \\
		
		\hline
	\end{tabular}
\end{table}

	\begin{table}[h] \centering \caption{Key parameters for Cluster 2 resulting from the UMAP-based clustering methodology, and containing GRBs with potential off-axis characteristics.} 
		\begin{tabular}{lccc}
			\hline
			\hline
			& fluence & E$_{peak}$ &  T$_{90}$\\ 
			& $(10^{-6}$ \textnormal{erg/cm$^{2})$} & \textnormal{(MeV)} & (\textnormal{s})\\
			
			\hline
			lower limit & 0.005 & 0.007 & 0.008\\
			upper limit & 11.1 $\times$10$^{-5}$ & 1.17 $\times$10$^{14}$&90.11\\
			centroid & 0.40 & 596.53 & 0.79 \\
			\hline
			\label{cluster1}
		\end{tabular} 
	\end{table}

	Following the identification of Cluster 2, an additional filtering step was applied to its elements in order to further refine the selection of candidates. The new constraint required that only GRBs with an error radius of no more than 0.1 degrees were considered. This threshold was selected to enhance the precision of localization, ensuring that the GRB possess well defined positions, which is a crucial factor when searching for optical counterparts or host galaxies. 
	It is important to note that we did not apply this filter at the beginning of the clustering process in order to avoid the introduction of a bias. The error radius may be influenced by the overlap of data from other facilites, which may have measured the redshift of detected the X-ray afterglow. By applying this filter only after the clustering, we aim to ensure a comprehensive analysis of the dataset, without excluding any potential candidate. 
	The error radius filter reduced the Cluster 2 dimension from 657 to 75 GRBs. Subsequently, a detailed characterization of these candidates was conducted, and the “gold” sample included those events that not only displayed an X-ray afterglow but also exhibited reliable redshift measurement. This procedure led to the identification of a sample of nine “gold” events, including our references GRBs (Table \ref{candidates}).

	\section{Merger-associated GRB sample}
	\label{sec:merger-GRB}

	\begin{table*}[h!] \centering 
		\caption{Properties of the reference GRBs and the most promising candidates identified by the clustering procedure.  }  
		\begin{tabular}{lcccccccc}
			\hline
			\hline
			& t$_{90}^{a}$  & fluence$^{a}$ & E$_{peak}^{a}$  & t$_{90}^{b}$  & fluence$^{b}$ & E$_{peak}^{b}$& redshift & distance$^{e}$\\
			
			& (\textnormal{s}) &\textnormal{(10$^{-7}$ erg/cm$^{2}$)}  & (\textnormal{keV})&  (\textnormal{s}) &\textnormal{10$^{-7}$ erg/cm$^{2}$}& (\textnormal{ph/cm$^{2}$/sec})& & (\textnormal{Mpc})\\
			\hline
			080905A &0.96$\pm{0.35}$ &8.50$\pm{0.46}$ &317.17$\pm{52.54}$&1&1.4$\pm{0.2}$&1.3$\pm{0.2}$&0.122& 524.86\\
			090510&0.96$\pm{0.14}$&33.7$\pm{0.41}$&4248.14$\pm{440.05}$&0.3&3.4$\pm{0.4}$&9.7$\pm{1.1}$&0.903& 3147.53\\
			090927&0.512$\pm{0.231}$&	3.03$\pm{0.18}$&195.22 $\pm{69.05}$	&2.2&2.0$\pm{0.3}$&2.0$\pm{0.3}$&1.37 & 4224.11\\
			
			131004A&  1.15$\pm{0.59}$& 5.10$\pm{0.19}$& 118.12$\pm{24.42}$&1.54&2.8$\pm{0.2}$&3.4$\pm{0.2}$& 0.71&26608.86\\
			
			150101B & 0.48$\pm{0.10}$  & 0.76$\pm{0.11}$ &  208.18$\pm{109.55}$  &0.02&0.23$\pm{0.06}$&-& 0.134&574.32\\
			170817A& 2.48$\pm{0.47}$ &2.79$\pm{0.17}$ & 214.70$\pm{56.59}$& -  & -& - & 0.009&39.79\\
			160821B&1.09 $\pm{0.98}$ &1.95$\pm{0.2}$  &38.17 $\pm{27.48}$ &0.48&1.0$\pm{0.1}$&1.7$\pm{0.2}$&0.161&685.88\\
			191031D & $0.25 \pm{0.02}$ & $39.68 \pm{0.15}$ & $681.54 \pm{32.55}$& 0.29 & 4.1$\pm{0.4}$ & 4.3$\pm{0.4}$ &0.5$^{c}$&1946.42\\
			210323A & 0.96$\pm{0.78}$ & 10.92 $\pm{0.30}$& 1440.78$\pm{339.89}$&1.12 & 2.4$\pm{0.3}$& 2.9$\pm{0.3}$& 0.733$^{d}$&2676.40\\
			
			\hline
			
			\hline
			
		\end{tabular} 
		\label{candidates}
		\tablefoot{Measurements are indicated as follows: $^{a}$ for values obtained from  Fermi GBM, $^{b}$ for values from  Swift/BAT, $^{c}$ for the photometric redshift from \citep{o2022deep} and$^{d}$ for the photometric redshift from \cite{nugent2024population}. The distances, denoted by $^{e}$, are derived using the {\tt Planck18} cosmology module from {\tt astropy}.}
	\end{table*}
	
	In this section we discuss the “gold” sample selected through both clustering and filtering procedures, and which properties are illustrated in Table \ref{candidates}. Aside for the two known KN-associated events, four of the nine gold events emerged as a particularly promising subset due to distinctive characteristics that align closely with KN outbursts:
	
	GRB 080905A lacks any associated supernova (SN) but enabled the detection of a faint afterglow within its host galaxy at redshift z=0.122 \citep{guelbenzu2021vlt}, a large spiral structure. This burst was observed in a region of minimal star formation and at a considerable distance from the galactic center, making a massive star progenitor unlikely and aligning with environments typical for sGRBs  \citep{rowlinson2010discovery}. The optical counterpart appears to adhere to a standard afterglow model, although the steep early X-ray decay and absence of X-ray detections beyond 1,000 seconds indicate a constrained late-time afterglow, potentially allowing for faint KN contributions.
	
	GRB 090510A was an energetic event, detected also by  Fermi-LAT \citep{ackermann2010fermi}. It was precisely localized by the Nordic Optical Telescope \citep{olofsson2009grb}, thereby facilitating the immediate association with its host galaxy, which was determined to be a late-type galaxy at z=0.903 \citep{2009GCN..9353....1R}. The burst exhibited key characteristics, including duration, environment, and spectral hardness, which are consistent with an origin in a compact-object merger \citep{ghirlanda2010onset,ruffini2016grb}.  Swift XRT observations (0.3–10 keV) revealed the presence of non-thermal X-ray emission, a weak precursor signal, and GeV-range emission. These distinct characteristics also align with the expected signatures of a KN-powered burst.

	GRB 160821B exhibited one of the best-sampled afterglows among sGRBs, as evidenced by a comprehensive multi-wavelength dataset that includes X-ray, optical, and radio observations \citep{lamb2019short}. The KN component, identified by excess red and nIR emission in the days following the burst, is consistent with models of neutron star mergers ejecting neutron-rich material. The burst’s environment and characteristics support a compact binary merger origin, with observations consistent with those of other sGRBs associated with KN emissions \citep{troja2019afterglow}.
	
	GRB 210323A was a short-hard burst. The event exhibited a long-lasting X-ray plateau, lasting approximately 10 $^{4}$ s, followed by a rapid decay. While an optical counterpart was not detected, \citep{shan2024grb} proposed that a supra-massive magnetar, formed after the merger of two neutron stars, could be the central engine responsible for the event. The extended plateau emission, combined with the subsequent sharp decline, is consistent with models involving energy injection from a supra-massive neutron star that eventually collapses into a black hole.

	Our gold sample includes three others GRBs: 090927,  131004A and  191031D. These GRBs lack a confirmed SN associations and their classification remains uncertain, although a collapsar progenitor has been considered for GRBs 090927 \citep{guelbenzu2012multi}. However, limitations in current data prevent us from drawing firm conclusions about their origins.

	\section{Host Association pipeline for nearby GRB events}
	\label{sec:hostpipeline}
	
	Our clustering approach allows us to select sGRBs detected by  Fermi potentially linked to a KN and aid follow-up observations. To further enhance the search of the electromagnetic counterpart, we developed a pipeline to identify potential host galaxies, assuming the sGRB in consideration occurs within a relatively nearby region of the universe, as it was for GRB 170817A.
	
	The pipeline utilizes the NASA/IPAC Extragalactic Database (NED) Local Volume Sample (LVS), chosen for its extensive completeness up to a distance of 80 Mpc \citep{cook2023completeness}, which outperforms other databases like GLADE+ \citep{dalya2018glade} or HECATE \citep{kovlakas2021heraklion} in terms of completeness within this range. Beyond this, NED remains robust, achieving 70$\%$ completeness out to roughly 300 Mpc. This ensures that the majority of potential host galaxies in the local universe are included, maximizing the reliability of our candidate selection. The filtering process begins by selecting galaxies within a 1 square degree search region around the Fermi GRB position. This search area is chosen to narrow down the number of galaxies and reduce the probability of false associations with unrelated objects.
	
\begin{figure}[h]
		\centering
		\includegraphics[width=\hsize]{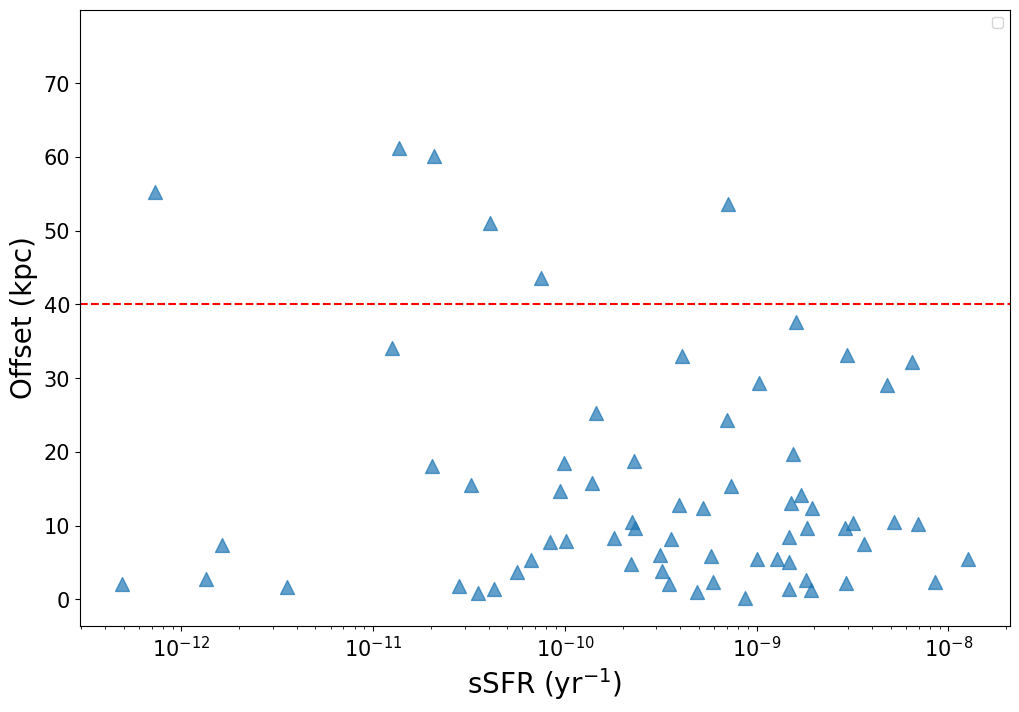}
		\caption{ Distribution of the sSFRs as a function of the physical offsets. Data from the catalog by \citet{nugent2024population}.}
		\label{ssfr}
	\end{figure}

	We also considered a physical offset constraint for the association GRB-host galaxy, limiting their offset to values lower than 40 kpc. This threshold is consistent with typical sGRB offsets observed in the literature, reflecting the tendency of these bursts to occur at significant distances from their host galaxy centers, likely due to natal kicks during binary neutron star formation \citep{fong2013locations, fong2022short, nugent2024population}. 
	Finally, a flag is applied based on the specific star formation rate (sSFR), highlighting galaxies within a range from  10$^{-12}$yr$^{-1}$ to 10$^{-8}$yr$^{-1}$ (illustrated in Figure \ref{ssfr}). To determine this range, we analyzed sGRBs features from the catalog by \citet{nugent2024population}, deriving their sSFR values. This interval is consistent with the association of sGRBs with environments that exhibit low star formation activity, such as elliptical or lenticular galaxies \citep{berger2014short}. It is important to stress that this is not a filtering parameter; rather, it highlights the nature of potentially selected galaxies with older stellar populations within the broader set of candidates identified by our pipeline.
	As a demonstration of the pipeline’s effectiveness, we tested it on GRB 170817A, the only event in the "gold" sample close enough to fall within the limits of the NED LVS catalog (see Table \ref{candidates}). The pipeline successfully identified its known host galaxy, NGC 4993, further validating the utility of our approach for nearby events. 
	
	To assess the utility of the pipeline, we performed a retroactive search for the 66  Fermi GRBs of Cluster 2 not in the “gold” sample and for which no redshift was measured. For this purpose, we referenced the Transient Name Server (TNS) catalog\footnote{https://www.wis-tns.org} in order to identify potential optical counterparts that are close to the host galaxies identified using the above pipeline. Although the TNS catalog is limited to GRBs discovered since 2014, it provides a sufficient level of statistical significance to validate our method, as it contains extensive observations of optical transients over nearly a decade.
	For GRBs with known afterglows, we performed a search within a 10-arcsec radius of the X-ray position, targeting the immediate vicinity for any associated optical counterparts. For GRBs without detected afterglows, candidate host galaxies were first identified using the NED LVS database, and then the search was performed in TNS. Additionally, a filter was applied to the discovery time of the potential  optical counterparts, selecting only events detected within a range of three months from the burst. Our search identified one candidate as a potential counterpart of GRB 220412713. We selected 82 possible host galaxies within 1 square degree of the Fermi GRB position. In a crossmatch with the TNS catalog, we identified AT 2022kpe at 3.68 arcseconds from WISEA J212432.06-000919.7, located at approximately 36 kpc from the GRB, suggesting a possible positional association. However, this transient was discovered by ZTF on May 19, one month after the GRB detection, as a rising source, suggesting that this event is unrelated with GRB 220412713. 
	
	The methodology we propose holds significant potential for future applications, particularly in KN event searches. By systematically identifying and prioritizing likely host galaxies, combined with the ability to pinpoint a GRB class with properties similar to GRB 170817A and GRB 150101B, our pipeline offers a valuable tool for enhancing targeted follow-up observations as GRB catalogs and GW detections expand. While currently limited to nearby events, this approach could be particularly beneficial for telescopes without wide-field capabilities, simplifying observational campaigns for optical counterparts. Furthermore, as facilities like the Vera C. Rubin Observatory’s Legacy Survey of Space and Time (LSST)\citep{2019ApJ...873..111I} become operational, their wide-field observations could complement our pipeline. The LSST’s ability to survey large sections of the Fermi error box and identify numerous galaxies will allow for efficient crossmatching with our results, further improving follow-up efforts.

	\section{Discussion}
	\label{discussion}
	Our methodology effectively identified a promising cluster of GRBs that exhibit characteristics consistent with potential sGRB-merger associations. The results align with previous studies that have found similar structures within GRB populations. In particular, the selection of $n_{cluster}$=5 is further supported by prior studies that consistently identified five clusters as optimal, despite variations in datasets, parameter choices, and clustering methodologies. For instance, \citet{acuner2018clustering} analyzed the Fermi GBM dataset, incorporating spectral parameters such as E$_{peak}$, fluence, and T$_{90}$, and employed Gaussian Mixture Models (GMM) for clustering. Their analysis highlighted the presence of five distinct groups, which were interpreted as representing different emission mechanisms, specifically distinguishing between photospheric and synchrotron origins during the GRB prompt phase. Similarly, \citet{Dimple:2023wvs} and \citet{misra2024diversity} explored clustering using the  Swift/BAT and Fermi GBM datasets, respectively, employing combinations of dimensionality reduction techniques such as PCA and t-SNE. They also included both spectral and temporal parameters in their analyses, enabling a detailed characterization of the GRB population. Both studies reported a consistent five-cluster solution, with two clusters identified as KN-associated groups based on the inclusion of known KN-associated GRBs. Furthermore, \citet{chattopadhyay2017gaussian} applied K-Means and Gaussian Mixture Models to the BATSE dataset and reported five clusters, attributing the separation to differences in burst energetics and durations. These findings underscore the robustness and potential universality of the five-cluster framework across diverse instruments, energy ranges, and parameter selections, establishing it as a valuable tool for exploring GRB subpopulations and their properties.
	
	To further validate our selection methodology, we compared our results with the sample of KN candidates identified by \citet{troja2023eighteen} in their comprehensive review of  Swift-detected GRBs. Swift, launched in 2004, is a multi-wavelength instrument equipped with the Burst Alert Telescope (BAT) \citep{barthelmy2005burst}, which operates in the 15–150 keV energy range, focusing on the detection of GRBs in the lower-energy regime. In contrast, Fermi-GBM covers a broader energy range, from 8 keV up to 40 MeV, making it more sensitive to high-energy gamma-ray emissions. Swift's additional instruments, the X-Ray Telescope (XRT) \citep{burrows2005swift} and the Ultraviolet/Optical Telescope (UVOT) \citep{roming2005swift}, enable precise afterglow localization and multi-wavelength follow-up observations. Together, these features make Swift a critical tool for identifying GRBs and their associated counterparts.
	
	We found that all three GRB events classified by \citet{troja2023eighteen} as ideal KN candidates (referred to as Group 1) and that have been observed by Fermi are in our cluster. These candidates are characterized by robust host galaxy associations, reliable redshift measurements, and an optical/near-infrared (nIR) excess consistent with KN emission above the non-thermal afterglow. The \citet{troja2023eighteen} Group 1 contains nine GRBs and the candidates not included in our analysis were either undetected by Fermi or occurred before its operational period. The lack of detection of some Swift GRBs reported by \citet{troja2023eighteen} can be attributed to Fermi's sensitivity to higher energies, further supporting the hypothesis that off-axis GRBs are predominantly soft events \citep{band2003comparison}. Moreover, identifying the same candidates using Fermi data, derived from an instrument with distinct energy coverage and detection capabilities, underscores the robustness of our clustering method. This highlights the complementarity of  Swift and Fermi: while Swift excels in the precision localization and follow-up of GRBs, Fermi's sensitivity to a broader energy range captures additional details of the gamma-ray emission.
	
	It is also worth noting that the only two long events, namely GRB 211211A \citep{rastinejad2022kilonova,troja2022nearby} and GRB 230307A \citep{levan2024jwst, dichiara2023luminous}, associated with KN events and detected by Fermi, did not fall within our cluster. These GRBs have durations that significantly exceed 2 seconds, distinguishing them from other GRBs typically associated with KNe. Their absence from our cluster suggests that the underlying mechanisms behind these events may be more complex than previously understood, or that they are on-axis events. Interestingly, we found that these two events were part of a separate cluster in our analysis, Cluster 1, which properties are summarized in Table \ref{clusterlong}. While our study primarily focuses on sGRBs, the observation of GRB 211211A and GRB 230307A challenge the traditional short-long classification paradigm and may require a separate discussion. However, long GRBs are generally dominated by events like broad-lined Type Ic SNe \citep{galama1999possible, hjorth2003very}, and their population also includes a large number of low-luminosity, soft GRBs whose peak emission is often below the energy range interval of Fermi GBM. Including long GRBs without clear selection criteria would introduce a strong observational bias due to  Fermi's reduced sensitivity to low-energy transients. A more robust investigation of potential long-duration merger-driven GRBs requires more sensitive instruments at soft X-rays, such as SVOM \citet{bernardini2021svom,atteia2022svom}, Einstein Probe \citet{yuan2022einstein}, or  THESEUS \citet{amati2018theseus}, and a significantly larger sample. 
	
	\begin{table}[h] 
		\centering 
		\caption{Key parameters for Cluster 1 resulting from the UMAP-based clustering methodology.} 
		\begin{tabular}{lccc}
			\hline
			\hline
			& Fluence & E$_{peak}$ &  T$_{90}$\\ 
			& ($10^{-6}$ \textnormal{erg/cm$^{2})$} & \textnormal{(keV)} & \textnormal{(s)}\\
			\hline
			lower limit & 3.59 & 38.10 & 2.69\\
			upper limit & 3147.5 & 1936.72 & 478.24\\
			centroid & 19.75 & 186.60 & 51.72 \\
			\hline
			\label{clusterlong}
		\end{tabular} 
	\end{table}
	
	Still, the presence of these two events within the same subpopulation hints a potential similarity in their origin and highlights the need for further research into the physical processes driving these bursts and their relationship with KNe. This finding aligns with the results of \citet{misra2024diversity}, who also identified a close proximity between these two events within their clustering analysis. This evidence also suggests the possibility of a similar progenitor, further supporting the hypothesis of a second distinct subpopulation potentially associated with KNe. Additional investigations into this cluster are in progress. 
	
	\section{Conclusion}
	\label{sec:conclusion}
	The confirmation of the connection between sGRBs and compact object mergers by groundbreaking events such as GRB 170817A has highlighted the necessity for systematic methods for identifying similar events when a simultaneous GW detection is missing. Building on this framework, our study developed a clustering methodology and an integrated pipeline to isolate GRBs with merger-like properties and select potential host galaxies, tailored to analyze the Fermi GBM bursts. 
	
	\begin{enumerate}
		\item The clustering methodology identified a subset of GRBs with properties consistent with known sGRBs and off-axis merger events: GRB 170817A and GRB 150101B. This approach demonstrates the potential of unsupervised learning techniques, such as K-Means, to isolate astrophysical events with specific characteristics, even within large and diverse datasets. Evaluation of clustering quality and consistency across dimensionality reduction techniques validated the robustness of the identified clusters. 
		\item A significant number of sGRBs in our selected sample with known redshifts exhibit characteristics indicative of merger-like origins, such as low star formation rates and significant physical offsets from galaxy centers.  The selected sample in this study is constrained to an error radius of no more than 0.1 degrees to ensure better localization.  The complete Cluster 2 includes a larger number of candidates, which is currently undergoing further analysis.
		\item We developed a pipeline that identifies potential host galaxies for sGRBs using the NED LVS. The pipeline successfully identified the host galaxy of GRB 170817A, confirming its effectiveness for nearby events and demonstrating its potential for future KN-associated GRB searches. Additionally, we tested it through a retroactive search on the GRBs within our Cluster 2 but without redshift measurements, although no new optical counterparts were found. The pipeline also has strong potential to complement current and upcoming facilities, such as the Vera C. Rubin Observatory's LSST and gravitational wave detectors like LIGO, Virgo, and KAGRA, by streamlining follow-up observations of sGRBs likely associated with KNe. 
		\item Results from previous studies support our cluster number selection. Additionally, a comparison with the sGRB and KN candidate sample from \citet{troja2023eighteen} reinforces our approach, as we successfully recover most of their Group 1 candidates.
		\item Overall, the clustering and filtering pipeline provides a scalable framework for future KN searches, facilitating the rapid identification of sGRBs with merger-like properties.  This  methodology allows for the classification of newly detected GRBs in few minutes, offering a time-efficient tool for real-time follow-up prioritization. As sGRB and GW catalogs expand with increasing detector sensitivity, this method can significantly improve the efficiency of multimessenger follow-up observations.

	\end{enumerate}

	Moreover, this approach can complement the capabilities of existing and forthcoming GW detectors, such as LIGO \citep{aasi2015advanced}, Virgo \citep{acernese2014advanced}, and KAGRA \citep{aso2013interferometer}, by focusing follow-up observations on the $\gamma$-ray events most likely to be associated with mergers and host environments for sGRBs. As detector sensitivity improves with the Einstein Telescope for GWs \citep{punturo2010einstein} and new facilities for GRB detection, such as Einstein Probe \citep{yuan2022einstein} and SVOM (\citet{bernardini2021svom}; \citet{atteia2022svom}), the probability of detecting nearby sGRB-GW associations will only increase. Looking forward, our methodology can be expanded to analyze both GW-associated and non-GW-associated GRBs, offering the potential to uncover a broader population of merger-related events. Additionally, integrating joint temporal studies and spectroscopic analyses \citep{negro2025prompt, camisasca2023grb} will further refine candidate identification, enhancing our ability to probe the physical properties of these bursts and advancing our understanding of compact object mergers.
	
	\emph{Software}:
	{\tt pandas} \citep{team2020pandas}, {\tt matplotlib} \citep{hunter2007matplotlib}, {\tt numpy} \citep{harris2020array}, {\tt scikit-learn} \citep{pedregosa2011scikit}, {\tt UMAP} \citep{mcinnes2018umap}.

	\begin{acknowledgements}
		LI acknowledges financial support from the YES Data Grant Program (PI: Izzo) Multi-wavelength and multi messenger analysis of relativistic supernovae. IFG acknowledges financial support from Piano Nazionale di Ripresa e Resilienza (PNRR) through the ETIC – Einstein Telescope Infrastructure Consortium (Missione 4/ Componente 2/ Investimento 3.1 - IR0000004). EC acknowledges support from MIUR, PRIN 2020 METE (grant 2020KB33TP).
		
	\end{acknowledgements}

	\bibliographystyle{aa} 

	\begin{appendix}
		\section{Clustering without dimensionality reduction}
		\label{sec:nored}
\begin{figure}[!ht]
			\begin{minipage}[t]{0.5\textwidth}
				\centering
				\includegraphics[width=\textwidth]{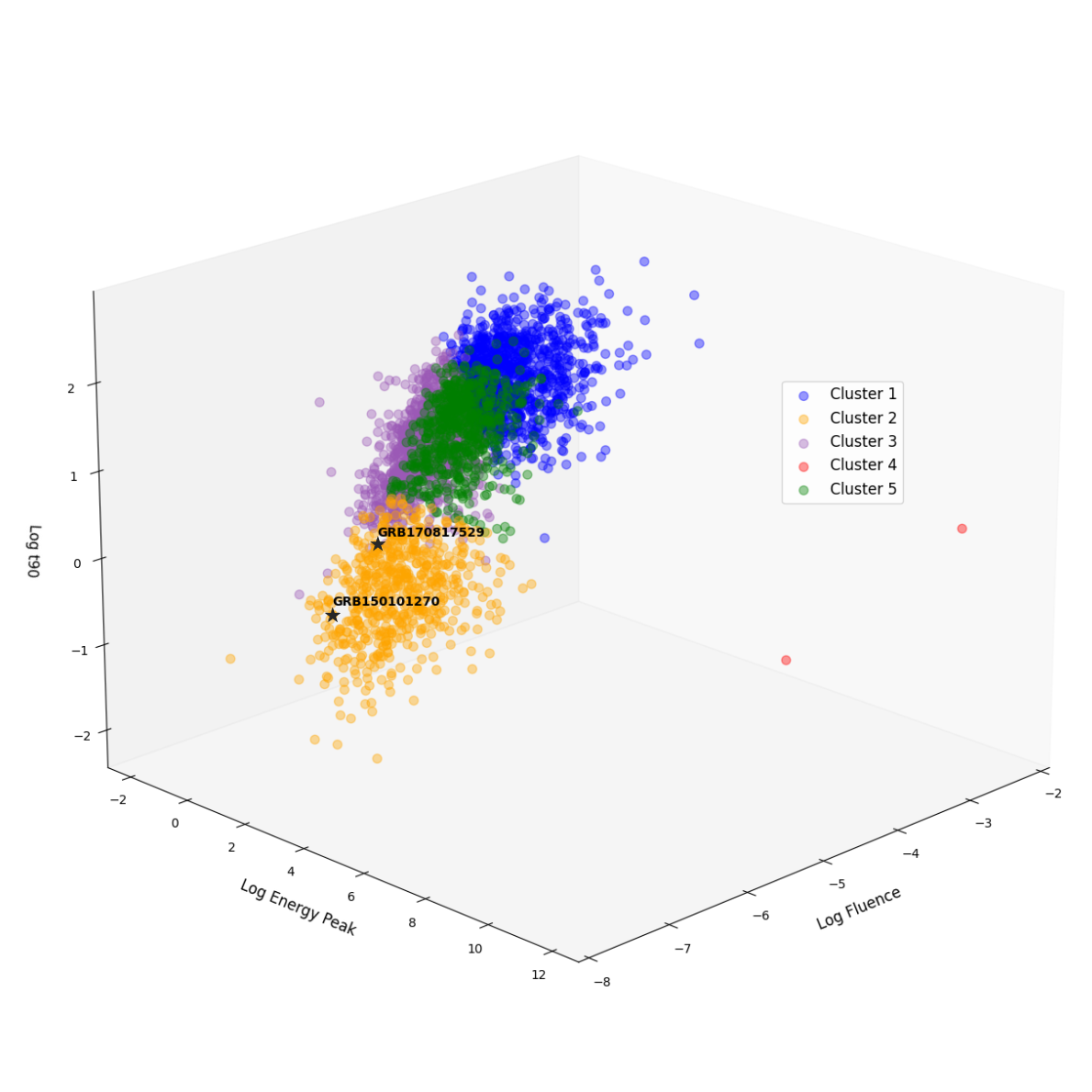}
				\caption{K-Means clustering results with $n_{cluster}$ = 5, without the application of dimensionality reduction techniques. }
				\label{fig:nored}
			\end{minipage}
		\end{figure}

		In order to ensure a comprehensive analysis, the dataset was tested without dimensionality reduction, with the application of only a logarithmic transformation and standardization. As in the previous case, a range of $n_{cluster}$ from 2 to 8 was analyzed. In each case, the two KN-associated GRBs were consistently clustered together, as was observed with the dimensionality-reduced data (see Figure \ref{fig:nored}). However, the clustering goodness remained poor across all values of $n_{cluster}$, suggesting that dimensionality reduction is necessary. In particular, we observe a DBI score close to 1, indicating poor clustering performance, along with a low Silhouette score, further confirming weak cluster separation (see Table \ref{tab:comparison_nored}). 
	These low metric values imply not only significant contamination between clusters, but also a lack of compactness and separability, and a high level of noise masking any underlying structure. Therefore, we excluded the non-reduced case from further consideration. The dimensionality reduction step is essential for improving the quality and interpretability of the clustering, as it helps highlight the most relevant components of the dataset and mitigate the impact of noisy or redundant information.
		
		\begin{table}[h!]
			\centering
			\caption{Clustering score without the application of the dimensionality reduction for $n_{cluster}$=4 and 5.}
			\label{tab:comparison_nored}
			\begin{tabular}{lcc} 
				\hline
				\hline
				$n\_cluster$ & DBI & Silhouette   \\
				\hline
				4 & 0.86& 0.31 \\
				5 &0.90 & 0.29 \\
				\hline
			\end{tabular}
		\end{table}
		
		\section{DBSCAN comparison}
		\label{sec:dbscan}
		
		To further validate our choice of $n_{cluster}$=5 we evaluated the consistency of the clustering results across different algorithms. We compared the degree of agreement between K-Means and an alternative clustering method: DBSCAN, both applied on the three dimensionality reduction techniques. We employed the {\tt DBSCAN} module from {\tt sklearn.cluster}. In this case, the ARI and NMI scores for $n_{cluster}$ = 5 were consistently higher as shown in Table \ref{tab:comparisonarinmi}). 
		This comparison reinforces our choice of $n_{cluster}$ = 5, as it demonstrates that this clustering configuration remains stable and consistent, irrespective of the method used.

		\begin{table}[h!]
			\centering
			\caption{Comparison of clustering performance scores between DBSCAN and K-Means for $n_{cluster}$ = 4 and 5, across dimensionality reduction techniques PCA, t-SNE, and UMAP.}
			\label{tab:comparisonarinmi}
			\begin{tabular}{lccc} 
				\hline
				\hline
				Method & ARI & NMI & $n\_cluster$ \\
				\hline
				
				t-SNE & 0.06 & 0.43 & 4 \\
				t-SNE & 0.09 & 0.48 & 5 \\
				\hline
				PCA & 0.002 & 0.008 & 4 \\
				PCA & 0.0008 & 0.006 & 5 \\
				\hline
				UMAP & 0.05 & 0.39 & 4 \\
				UMAP & 0.06 & 0.44 & 5 \\
				\hline
			\end{tabular}
		\end{table}
	\clearpage
	\onecolumn
	\section{Cluster distributions}

		\label{distr}
	\begin{figure*}[ht!]
		\centering
		
		\begin{subfigure}{1\textwidth}
			\centering
			\includegraphics[width=\textwidth]{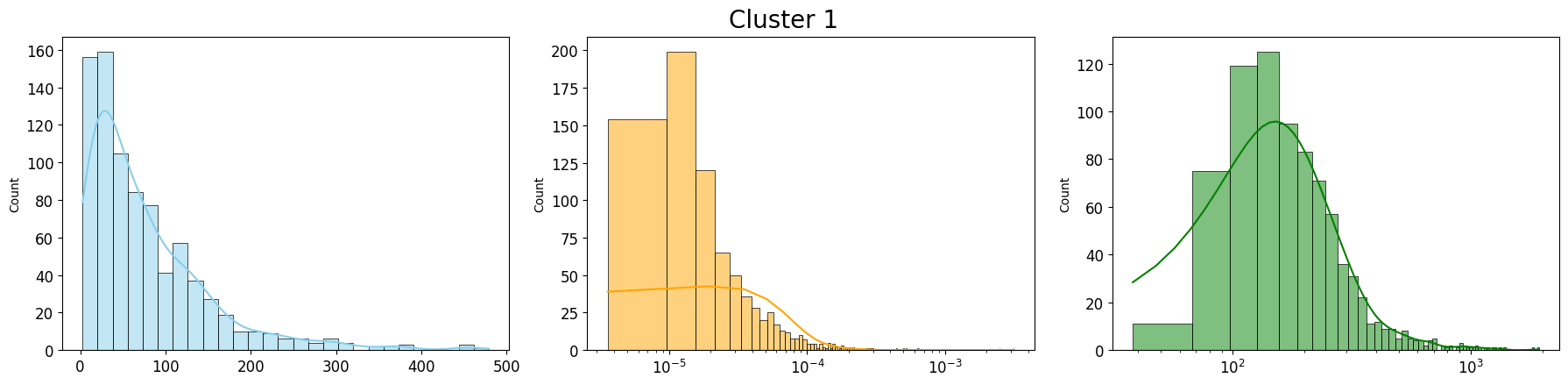}
		\end{subfigure}

		\begin{subfigure}{1\textwidth}
		\centering
		\includegraphics[width=\textwidth]{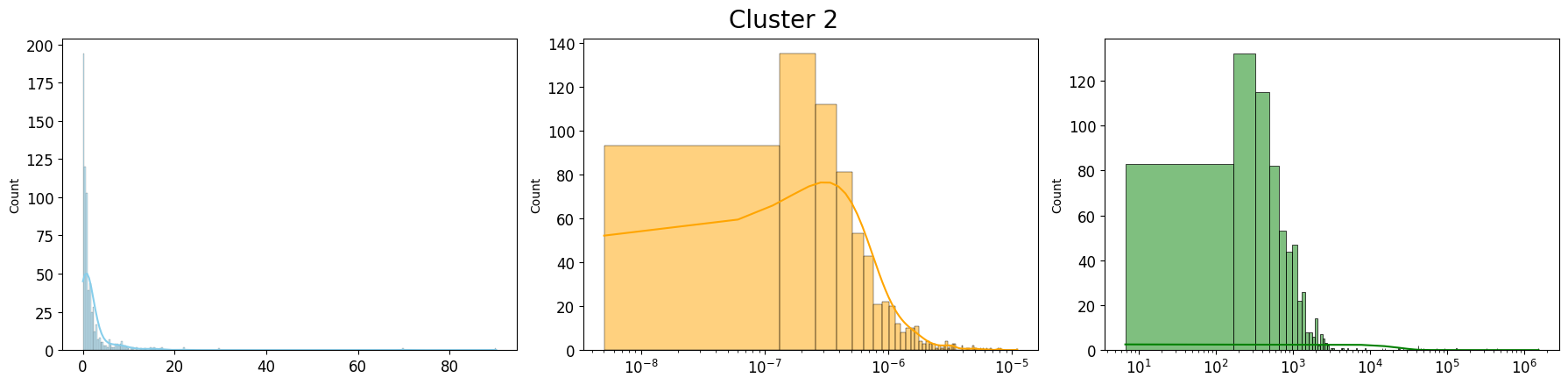}
	\end{subfigure}

		\begin{subfigure}{1\textwidth}
		\centering
		\includegraphics[width=\textwidth]{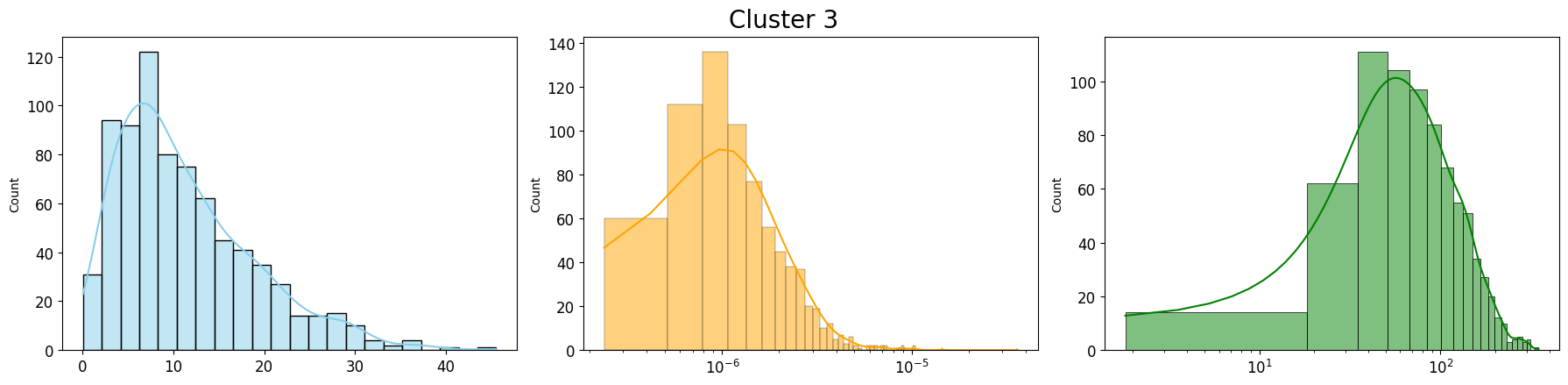}
	\end{subfigure}
	\begin{subfigure}{1\textwidth}
		\centering
		\includegraphics[width=\textwidth]{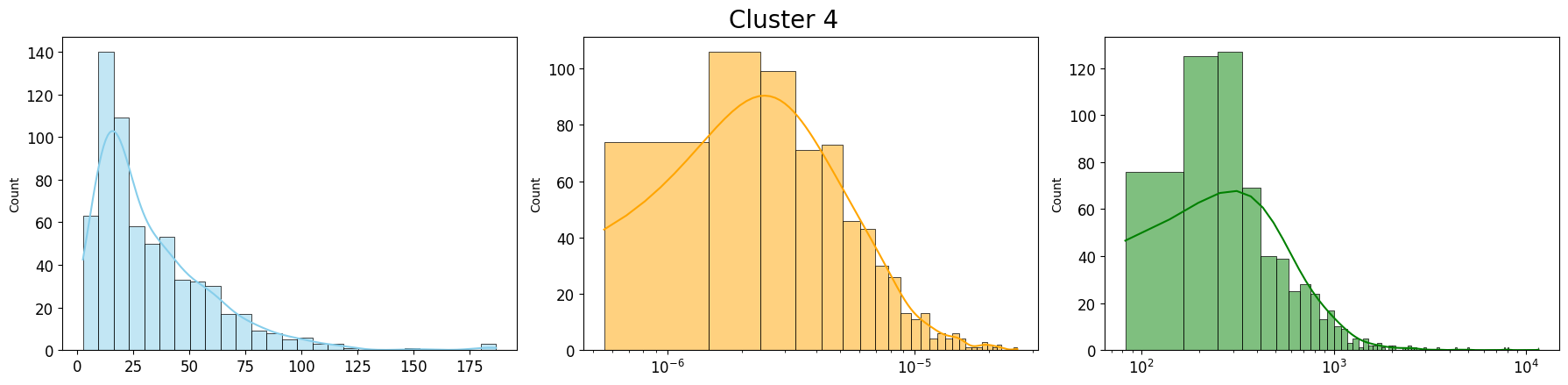}
	\end{subfigure}
	
	\begin{subfigure}{1\textwidth}
		\centering
		\includegraphics[width=\textwidth]{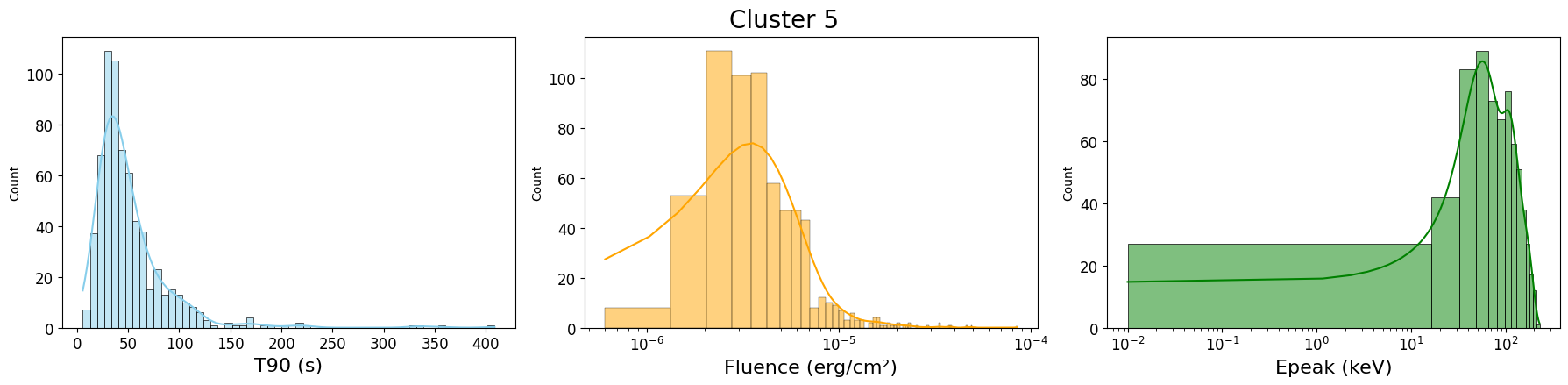}
	\end{subfigure}
		
		\caption{Distributions of T$_{90}$, fluence, and E$_{peak}$ for each cluster obtained using UMAP.}
		\label{fig:umapdistributions}
	\end{figure*}
	\clearpage
	\onecolumn
		\section{GRBs Sample} 
		\begin{longtable}{c  ccc}
			\caption{Cluster 2 GRBs selected through UMAP dimensionality reduction and refined by filtering based on error radius.} \\
			\hline
			\hline
			name         &           T$_{90}$&      fluence &   E$_{peak}$ \\
			& ($s$) & $\times10^{-7}$($erg/cm^{2}$) & ($keV$) \\
			\hline
			\endfirsthead
			\caption{continued.}\\
			\hline
			name         &           T$_{90}$ &      fluence &   E$_{peak}$\\
			& ($s$) & $\times10^{-7}$($erg/cm^{2}$) & ($keV$) \\
			\hline
			\endhead
			\hline
			\endfoot

			GRB090927422 &  $0.51 \pm{0.23}$ & $3.03 \pm{0.18}$ & $195.22 \pm{69.05}$ \\
			GRB220730659 &   $0.16 \pm{0.29}$ & $9.36 \pm{0.29}$ & $327.41 \pm{24.95}$ \\
			GRB081226044 &  $0.83 \pm{1.03}$ &  $4.30 \pm{0.23}$ & $584.59 \pm{284.04}$ \\
			GRB180727594 &  $0.89 \pm{0.29}$ & $3.94 \pm{0.16}$ & $56.05 \pm{3.37}$ \\
			GRB210119121 &  $0.05 \pm{0.06}$ & $1.87 \pm{0.14}$ &  $223.98 \pm{39.93}$ \\
			GRB200219317 &  $1.15 \pm{1.032}$ & $15.85 \pm{0.40}$ & $1507.49 \pm{204.90}$ \\
			GRB161001045 &   $2.24 \pm{0.23}$ & $17.63 \pm{0.15}$ & $372.61 \pm{59.60}$ \\
			GRB130515056 &  $0.26 \pm{0.09}$ & $10.89 \pm{0.16}$ & $452.86 \pm{74.42}$ \\
			GRB200411187 &   $1.44 \pm{0.51}$ & $6.58 \pm{0.23}$ & $361.20 \pm{77.71}$ \\
			GRB150301045 &  $0.42 \pm{0.29}$ & $1.81 \pm{0.12}$ & $149.42 \pm{77.77}$ \\
			GRB081226509 &  $0.19 \pm{0.14}$ & $3.44 \pm{0.27}$ & $359.82 \pm{77.08}$ \\
			GRB141102536 &  $2.62 \pm{0.33}$ & $14.19 \pm{0.41}$ & $598.64 \pm{92.24}$ \\
			GRB230116374 &  $0.06 \pm{0.04}$ & $29.53 \pm{0.28}$ & $573.82 \pm{30.18}$ \\
			GRB110420946 &  $0.13 \pm{0.52}$ & $2.44 \pm{0.27}$ & $353.52 \pm{87.07}$ \\
			GRB160411062 &  $0.67 \pm{0.43}$ & $2.25 \pm{0.19}$ & $161.53 \pm{41.37}$ \\
			GRB100625773 &   $0.24 \pm{0.27}$ & $5.63 \pm{0.24}$ &  $482.13 \pm{61.94}$ \\
			GRB090531775 &  $0.77 \pm{0.23}$ & $3.18\pm{0.18}$ & $1659.39 \pm{399.38}$ \\
			GRB200325138 &   $0.96 \pm{0.14}$ & $14.80 \pm{0.46}$ & $988.82 \pm{143.55}$ \\
			GRB170127634 &  $1.73 \pm{1.35}$ & $3.05 \pm{0.13}$ & $481.68 \pm{131.52}$ \\
			GRB120830297 &  $0.89 \pm{0.23}$ & $30.69 \pm{0.27}$ & $1088.52 \pm{105.29}$ \\
			GRB090510016 &   $0.96 \pm{0.14}$ & $33.732 \pm{0.41}$ & $4248.14 \pm{440.05}$ \\
			GRB120817168 &   $0.16 \pm{0.11}$ & $17.86 \pm{0.10}$ & $1348.83 \pm{217.61}$ \\
			GRB160612842 &  $0.29 \pm{0.23}$ &  $9.09 \pm{0.15}$ & $1263.31 \pm{394.01}$ \\
			GRB170817529 &  $2.048 \pm{0.47}$ & $2.79 \pm{0.17}$ & $214.70 \pm{56.59}$ \\
			GRB080905499 &   $0.96 \pm{0.35}$ & $8.50 \pm{0.46}$ & $317.17 \pm{52.54}$ \\
			GRB151228129 &  $0.26 \pm{2.27}$ & $4.63\pm{0.36}$ & $873.74 \pm{186.62}$ \\
			GRB180805543 &    $0.96 \pm{0.59}$ &  $3.66 \pm{0.12}$ & $393.43 \pm{58.64}$ \\
			GRB180402406 &  $0.445 \pm{0.33}$ & $13.12 \pm{0.28}$ &  $1328.87 \pm{161.03}$ \\
			GRB200216380 &  $8.19 \pm{1.64}$ &  $8.70 \pm{0.54}$ & $380.78 \pm{119.52}$ \\
			GRB140320092 &  $2.30 \pm{1.53}$ & $1.016 \pm{0.10}$ & $831.59 \pm{3338.75}$ \\
			GRB150101641 &   $0.08 \pm{0.93}$ & $2.380\pm{0.15}$ & $28.67 \pm{6.73}$ \\
			GRB130626452 &  $1.73 \pm{0.77}$ & $2.34 \pm{0.08}$ & $347.31 \pm{210.42}$ \\
			GRB160726065 &  $0.77 \pm{0.34}$ &  $10.12 \pm{0.16}$ & $271.37 \pm{72.59}$ \\
			GRB131004904 &   $1.15 \pm{0.59}$ & $5.09 \pm{0.19}$ & $118.13 \pm{24.42}$ \\
			GRB100117879 &  $0.25\pm{0.83}$ & $4.23 \pm{0.69}$ & $327.22 \pm{52.92}$ \\
			GRB180715755 &  $0.70 \pm{0.28}$ &  $10.45\pm{0.22}$ &   $1079.89 \pm{143.60}$ \\
			GRB220412713 &   $0.16 \pm{0.07}$ & $4.32 \pm{0.07}$ & $391.85 \pm{68.363}$ \\
			GRB081101491 &  $0.13 \pm{0.09}$ &  $1.68 \pm{0.04}$ & $236.56 \pm{55.41}$ \\
			GRB180718082 &   $0.08 \pm{0.18}$ & $0.85 \pm{0.17}$ &  $89.74 \pm{9.78}$ \\
			GRB140402007 &    $0.32 \pm{0.63}$ & $2.84 \pm{0.29}$ & $1093.34 \pm{232.98}$ \\
			GRB170318644 &  $4.09 \pm{2.32}$ &  $5.08 \pm{0.52}$ & $891.68 \pm{395.28}$ \\
			GRB201214672 &  $0.45 \pm{0.18}$ & $0.95\pm{0.14}$ &  $797.15 \pm{303.28}$ \\
			GRB100206563 &  $0.17 \pm{0.07}$ &  $7.56 \pm{0.11}$ & $454.31 \pm{63.63}$ \\
			GRB130912358 &  $0.51 \pm{0.14}$ &  $7.00\pm{0.17}$ & $407.68 \pm{163.61}$ \\
			GRB220617772 &  $0.70 \pm{0.23}$ & $6.56 \pm{0.11}$ & $1947.47 \pm{329.55}$ \\
			GRB240109168 &  $0.25 \pm{0.85}$ &  $0.95 \pm{0.14}$ & $472.85\pm{262.33}$ \\
			GRB170325331 &  $0.57 \pm{0.33}$ & $2.57 \pm{0.12}$ & $434.19 \pm{134.51}$ \\
			GRB111117510 &  $0.43\pm{0.08}$ & $5.64 \pm{0.13}$ &  $502.85 \pm{111.18}$ \\
			GRB180418281 &   $2.56 \pm{0.20}$ & $5.89 \pm{0.09}$ & $1051.09 \pm{951.48}$ \\
			GRB210323918 &   $0.96 \pm{0.78}$ & $10.92 \pm{0.30}$ & $1440.78 \pm{339.89}$ \\
			GRB240720017 &  $1.79 \pm{0.20}$ & $23.88 \pm{32.29}$ & $864.77 \pm{75.09}$ \\
			GRB180204109 &  $1.15 \pm{0.09}$ & $17.46 \pm{0.13}$ & $814.81 \pm{164.93}$ \\
			GRB221024542 &  $8.44 \pm{1.55}$ & $7.75 \pm{0.33}$ & $840.69 \pm{324.04}$ \\
			GRB100216422 &  $0.19 \pm{0.14}$ & $3.87 \pm{0.14}$ &  $228.56 \pm{151.42}$ \\
			GRB150101270 &   $0.48 \pm{0.95}$ & $0.76 \pm{0.11}$ & $208.18 \pm{109.55}$ \\
			GRB101008697 &   $8.96 \pm{1.84}$ & $13.49 \pm{0.43}$ & $2045.71 \pm{1179.78}$ \\
			GRB141205337 &   $1.28 \pm{0.57}$ & $10.65\pm{1.14}$ & $499.94 \pm{175.83}$ \\
			GRB240615744 &  $0.10 \pm{0.05}$ & $15.20\pm{0.07}$ & $660.55 \pm{67.37}$ \\
			GRB190427190 &  $0.37 \pm{0.07}$ & $5.34 \pm{0.12}$ & $1260.11 \pm{607.95}$ \\
			GRB201221963 &  $0.14 \pm{0.06}$ & $6.35 \pm{0.23}$ & $73.49 \pm{2.17}$ \\
			GRB171007498 &  $3.46 \pm{0.34}$ & $3.03 \pm{0.10}$ & $757.53 \pm{1605.87}$ \\
			GRB221120895 &   $0.64 \pm{0.26}$ & $4.96 \pm{0.11}$ & $499.36 \pm{53.73}$ \\
			GRB101224227 &   $1.73 \pm{1.68}$ & $1.91 \pm{0.27}$ & $253.47 \pm{217.98}$ \\
			GRB191031891 &  $0.25 \pm{0.02}$ & $39.68 \pm{0.15}$ & $681.54 \pm{32.55}$ \\
			GRB200623138 &  $0.09 \pm{0.14}$ & $0.63 \pm{0.11}$ & $230.62 \pm{45.03}$ \\
			GRB210421455 & $11.00 \pm{2.67}$ & $9.73 \pm{0.26}$ &  $1629.56\pm{548.73}$ \\
			GRB081024245 &  $0.83 \pm{1.28}$ & $1.98 \pm{0.17}$ & $996.35 \pm{397.96}$ \\
			GRB160821937 &  $1.09 \pm{0.98}$ & $1.95\pm{0.20}$ & $38.17\pm{27.48}$ \\
			GRB130716442 &  $0.76 \pm{0.38}$ & $6.40 \pm{0.88}$ & $883.46 \pm{1120.96}$ \\
			GRB160714097 &   $0.32 \pm{0.36}$ & $1.11 \pm{0.15}$ & $61.71 \pm{52.11}$ \\
			GRB110112934 &  $2.30 \pm{2.54}$ &  $4.05\pm{0.27}$ & $561.55 \pm{147.75}$ \\
			GRB090621922 &  $0.38 \pm{1.03}$ & $4.75 \pm{0.19}$ & $486.22 \pm{120.54}$ \\
			GRB231115650 &  $0.03 \pm{0.04}$ & $4.75 \pm{0.07}$ &  $787.01 \pm{107.08}$ \\
			GRB160408268 &  $1.05 \pm{0.60}$ & $6.98 \pm{0.28}$ & $841.97 \pm{170.45}$ \\
			GRB180618030 &   $3.71 \pm{0.58}$ &  $12.29 \pm{0.14}$ & $2292.46 \pm{874.31}$ \\
			
			\hline
			
		\end{longtable}

	\end{appendix}

\end{document}